\documentstyle[aps,floats,preprint,rotating,epsfig]{revtex}
\tightenlines

\begin{document}
\draft
\preprint{\vbox{
\hbox{hep-ph/0209338}
\hbox{September 2002}}}
\title{Signatures of Right-Handed
Majorana neutrinos and gauge bosons in $e\gamma$ 
Collisions}
\author{Nikolai Romanenko}
\address{
Ottawa-Carleton Institute for Physics \\
Department of Physics, Carleton University, Ottawa, Canada K1S 5B6}

\maketitle

\begin{abstract}
The process 
  $e^- \, \gamma  \rightarrow e^+ \, W_R^- \, W_R^-$ 
is studied in the framework of the Left-Right symmetric model.
It is shown that this reaction 
and $e^- \, \gamma  \rightarrow l^+ \,
 W_R^- \, W_R^-$ for the arbitrary final lepton
are likely to be discovered for CLIC collider
option.
 For relatively light
doubly charged Higgs boson its mass does not have much influence on the
discovery potential, while for heavier values the probability of the 
reaction increases.
\end{abstract}

\section{Introduction}
 Majorana neutrino masses arise naturally in many extensions of the
Standard Model (SM) such as singlet Majorana mass model \cite{seesaw},
Higgs triplet model \cite{Gelmini} or Left-Right symmetric model (LRM) 
\cite{LR}. One of the main sources of their popularity comes from the so-called
See-Saw mechanism \cite{seesaw} where
 the left-handed neutrinos turn out to be  
light due to the corresponding right-handed neutrinos being heavy.
The LRM can easily include heavy right-handed Majorana neutrinos and
the  
See-Saw mechanism since it   has a heavy
mass scale determined by the $SU(2)_R$ symmetry breaking. In this case
the LRM contains a triplet Higgs field with hypercharge $Y=2$ \cite{LR}.

 The theory with massive Majorana neutrinos, especially with heavy ones,
allows a variety of processes violating lepton number conservation.
They provide beautiful signatures which can be tested at various collider options.
%
%
Estimates of the discovery probability for heavy
 Majorana neutrinos at the LHC were done in \cite{LHC} and
for electron-positron colliders in \cite{e+e-}.
An electron-electron collider provides
 specific and very useful
option for Majorana neutrino search,
the so called ``the inverse double-$\beta$ decay'' process.
Together with some associated processes, it was studied in \cite{ee}. 
For the electron-proton option of HERA, Majorana neutrino production
was studied in \cite{HERA}, for
  VLHC (Very Large Hadron Collider)
in \cite{VLHCep}, for electron-photon collider -- in \cite{Peressutti}.

 All these estimates should be combined with the restrictions obtained from the 
neutrinoless double $\beta$--decay \cite{Double-beta},
 the well-known low-energy lepton-number 
violating process. The resulting numbers 
essentially depend on the chirality of neutrinos
and on the bosonic sector of the model. For example, limitations from 
double $\beta$--decay in the framework of the Standard Model (SM)
bosonic sector yield an upper bound on the effective electron 
neutrino mass $<m_{\nu_e}> \le 0.2$ eV, while for the case of
LRM heavy right-handed neutrinos are allowed with the lower
bound depending on the mass of the right-handed gauge boson
\cite{Double-beta}.

In this article we will concentrate on the electron-photon collider option.
We will study the signatures for heavy right-handed Majorana neutrinos
in the process $e^- \, \gamma \rightarrow e^+ \, W_R^- \, W_R^-$. This process
is analogous to the one considered in \cite{Peressutti} for the left-handed
 gauge bosons.
 However, since the current limitations for $W_R$ masses are relatively strict
 and imply
$M_{W_R} \ge  700$ GeV \cite{PDG},
 the abovementioned process with two final state $W_R$'s 
is likely to be discovered only
at the collider  mashines with very high center of mass energies. The CERN
Linear Collider (CLIC) proposal \cite{CLIC} is anticipated to provide
$\sqrt{s}=$ 3, 5 and 8 TeV energies, appropriate for the  process chosen.
We assume the integrated luminosity of $L=$ 500 fb$^{-1}$.
Final state $W_R$ bosons are expected to decay mostly into quark jets
and hence to be identified through the quark jets
with an appropriate invariant mass. 
Due to presence of the final state positron, the process 
has no SM background. 

In the next section we will briefly discuss the main features of the LRM
and discuss some phenomenological constraints on the parameters of this model.

\section{The Model}

 In this section we give a brief description of the LRM. For more detailed 
reference one can refer to \cite{LR}.The gauge symmetry of the LRM extends the SM 
gauge group to $SU(2)_L \times SU(2)_R \times U(1)_{B-L}$. The fermionic sector
contains, in addition to SM particles, right-handed neutrinos: one specie for 
each generation. Quarks and leptons transform under the gauge group as follows:
\begin{equation} 
{ Q}_{iL}=
\left[ \begin{array}{c} u \nonumber \\ d
\nonumber \end{array} \right]_{iL}=({2}, 1,
\frac{1}{3});\: 
{ Q}_{iR}=
\left[ \begin{array}{c} u \nonumber \\ d
\nonumber \end{array} \right]_{iR}=(1, {2},
\frac{1}{3}) 
\end{equation} 
\begin{equation} 
{ \Psi}_{iL}=
\left[ \begin{array}{c} \nu \nonumber \\ e
\nonumber \end{array} \right]_{iL}=({2}, 1, -1);\:
{ \Psi}_{iR}=
\left[ \begin{array}{c}
\nu \nonumber \\
e \nonumber \end{array} \right]_{iR}=(1, {2}, -1),
\end{equation}
where $i$
is the flavour index, $L$ and $R$ denote left-handed and right-handed 
chirality, $Q$ and $\Psi $ stand for quark and lepton wave functions
respectively.
The gauge sector includes  right-handed 
gauge bosons $W_R$ and $Z_R$ in addition to SM gauge bosons.
The greatest extension has to be done for the scalar sector.
In order to supply quarks and leptons with masses one needs the Higgs bidoublet field
with the following quantum numbers:
$$\Phi=\left(
\begin{array}{ll}
\phi_1^0 & \phi_1^+ \nonumber \\
\phi_2^- & \phi_2^0
\end{array}
\right)=({2}, {2}^*,0),
$$
and with the following vacuum expectation value (VEV)
$$ <\Phi>= \frac{1}{\sqrt2}\left(
\begin{array}{ll}
k_1 & 0 \\
0   &k_2
\end{array} \right).
$$
 Besides this, another Higgs field with nonzero
$B-L$ quantum number is necessary in order to provide symmetry breaking
of $SU(2)_L \times SU(2)_R \times U(1)_{B-L}$ to the SM gauge group.
The most popular way to do it  also gives rise to  Majorana masses 
for neutrinos: this way is to introduce a Higgs triplet field
\newcommand{\matr}{\left( \begin{array}}
\newcommand{\ematr}{\end{array} \right)}  
\begin{equation}
 {\displaystyle
\Delta_R=\matr{cc}\Delta_R^+/ \sqrt{2}&\Delta_R^{++}\\
\Delta_R^0&-\Delta_R^+ /\sqrt2\ematr = (1,3,2)}
\end{equation}
with the vacuum expectation value:
\begin{equation}
\begin{array}{c}  {\displaystyle\langle\Delta_{R}\rangle
=\frac1{\sqrt{2}}\matr{cc}0&0\\v_{R}&0
\ematr.}
\end{array}
\end{equation}  

For an explicit (manifest) $L \leftrightarrow R$
symmetry, the corresponding left-handed
Higgs-Majoron field should also be introduced:
\begin{equation}
 {\displaystyle 
\Delta_L=\matr{cc}\Delta_L^+/ \sqrt{2}&\Delta_L^{++}\\
\Delta_L^0&-\Delta_L^+ /\sqrt2\ematr = (3,1,2)}
\end{equation}
with the vacuum expectation value:
\begin{equation}
\begin{array}{c}  {\displaystyle
\langle\Delta_{L}\rangle
=\frac1{\sqrt{2}}\matr{cc}0&0\\v_{L}&0
\ematr}
\end{array}.
\end{equation}  

The Yukawa interactions of the Higgs triplets with fermions in the model
read:

\begin{equation}
-{\cal L }_{Yuk}=i h_{R,ll'}\Psi_{lR}^TC\sigma_2\Delta_R\Psi_{l'R} + 
i h_{L,ll'}\Psi_{'L}^TC\sigma_2\Delta_L\Psi_{l'L}
\:\:\:+\:{\rm h.c.}, 
\label{Yuk}
\end{equation}
  where
$l,l'$ are flavour indices, these interactions yield Majorana mass 
to neutrinos and are relevant to the process  studied 
in this article. Since left-handed neutrinos are practically massless
one expects $v_L$ to be small, while the value of $v_R$ provides
natural scale for the right-handed neutrino masses.
 After ignoring possible  mixing between 
lepton families the masses of right-handed Majorana neutrinos are 
given by $m_{Ri}= \sqrt2 h_{R,ii} v_R$.

 For further considerations I will choose $v_L=k_2=0$, this 
condition being compatible with phenomenological limits 
\cite{GunionLR}. The gauge couplings for the left-handed
and right-handed gauge groups are set  equal,
$g_L=g_R $. 

Present phenomenological bounds on the triplet Yukawa couplings
$h_{ll'}$ were discussed in \cite{GKR}, they satisfy
$h/M_{\Delta^{++}}<0.44$ TeV$^{-1}$. The limit on the masses of right-handed 
gauge bosons is $M_{W_R} > 700$ GeV \cite{PDG}. 

\section{Calculations and Results}

 Nine Feynman diagrams are involved in the 
$e^- \, \gamma \rightarrow e^+ \, W_R \, W_R $ process, see Fig.1.
Diagrams 1-4 do not contain doubly charged Higgs field $\Delta^{--}$
and correspond to the diagrams considered in \cite{Peressutti},  with the
proper change of chiralities of particles. Diagrams 5-9 involve
 virtual $\Delta^{--}$.
Using the CALCUL technique \cite{CALCUL} we obtained
 the expressions for chirality amplitudes
presented in the Appendix. Alternatively, the expressions for square matrix elements
were obtained by means of the COMPHEP package \cite{COMPHEP}.
We have checked the consistency of our results with 
\cite{Peressutti}: if the value of the charged boson mass is set to 80 GeV,
diagrams with doubly charged Higgs boson are neglected
and the initial photon energy is fixed, the results are the same
as in Fig. 2  of \cite{Peressutti}.

 In contrast to \cite{Peressutti} we have used the backscattered
laser photon spectrum \cite{backlaser}
  for the initial state photon so that the
cross section of the process 
is described by convoluting the fixed photon energy cross section 
with this spectrum:
\begin{equation}
\sigma= \int dx  f_{\gamma/e}(x,\sqrt{s}/2) \;
\hat{\sigma}(e^- \gamma \to e^+ W_R^- W_R^-).
\end{equation}

 The following final state  cuts were applied:
$| \cos \theta_{1f}|< 0.9$, $E_{e^+} > 10 GeV$,
these cuts are based on the detector considerations.

 In Figure 2 the cross section of the process is shown as a function of
electron-positron collision energy $\sqrt{s_{ee}}$.
In all three paints of Fig. 2 the  mass of
the doubly charged Higgs boson is 
600 GeV, while the masses of the
right-handed gauge boson are
$M_{W_R}=$ 700, 1000, 1500 GeV correspondingly.
Each of the paints contains 4 curves for different values of
Majorana neutrino mass: $M_N=$ 5 TeV (solid line),
$M_N=$ 3 TeV (long-dashed line),
$M_N=$ 1 TeV (short-dashed line) and $M_N=$ 500 GeV (dash-dotted line).
Thresholds of the curves are due to the Majorana neutrino 
propagator pole:
$$
\Pi (p_1+p_2-p_4-p_5,M_N,\Gamma_N)= 
\frac{1}{s+M_W^2-M_N^2-2(p_1+p_2)(p_4+p_5)+iM_N \Gamma_N}.
$$

 Figure 3 shows the effects of Majorana neutrino and doubly 
charged Higgs widths in the propagators of the amplitudes.
In that figure the
right-handed boson mass is 1 TeV, $\sqrt{s_{ee}}=5$ TeV.
The dashed-dotted line represents the cross section of the process as a 
function of Majorana neutrino mass with the constant decay widths
$\Gamma_N=\Gamma_{\Delta^{++}}=10$ GeV. In general, the  decay width
for heavy Majorana neutrino $N \rightarrow  W^{\pm} \, l^{\mp}$
(where $l$ stands  for massless lepton)
is given by:
\begin{equation}
\Gamma_N=\frac{g^2}{(32 \pi M_N^3 M_W^2)}
(M_N^2-M_W^2)(M_N^4+M_N^2 \cdot M_W^2 - 2M_W^4)
\label{wN}
\end{equation}
(in the case when $M_{W_R}>M_N$ the above mentioned mode is closed
but there may exist decay modes
to left-handed bosons through some mixing effects).
As for the decay modes of the doubly charged Higgs boson
it is useful to take  two fermion decays 
($\Delta^{++} \rightarrow l^+ l^+$)
into account explicitly:
\begin{equation}
\Gamma_{\Delta^{++}\rightarrow l^+ l^+ }= \frac{1}{8 \pi} h^2_{ll} M_{\Delta^{++}}
\end{equation}
(here $h_{ll} $ stands for the Yukawa coupling to lepton $l$)
while leaving possible bosonic decay width as a
free phenomenological parameter
(for more detailed discussion see \cite{GKR}):
\begin{equation}
\Gamma_{\Delta^{++}}= \Gamma_b+\Gamma_f.
\label{wd}
\end{equation}
In the following calculations we have chosen $\Gamma_b=10$ GeV.
The solid line in Fig. 3 corresponds to the mass dependent
widths $\Gamma_N$ and $\Gamma_{\Delta^{++}}$ according
to eqs. (\ref{wN}) and (\ref{wd}). However, 
$\Delta^{++}$ width does not play
much role in the estimates of the cross section:
 the dashed line in Fig. 3
($\Gamma_N$ as in eq. (\ref{wN}), $\Gamma_{\Delta^{++}}=10$ GeV)
is completely invisible since it coincides with the solid one.
As in \cite{Peressutti} the curves reach their highest values
within a certain range of $M_N$ masses
(``peak-like'' behaviour), though they do not decrease  much
(as in \cite{Peressutti}) when $M_N$ is above this range,
the latter happens due to effects of the doubly-charged Higgs boson.
In general one can state that the increase of Majorana neutrino's width
decreases the cross section of the process in the mass range
when the cross section has ``peak-like'' behaviour and has not much influence
on the cross-section away from that  ``peak''.

 Figs. 4 and 5 represent the cross section as a function of mass 
of right-handed boson for the three CLIC options:
$\sqrt{s_{ee}}=$ 3, 5, 8 TeV. In Fig. 4 all the effects of
doubly charged Higgs boson are neglected (only diagrams 1-4 are taken
into account, or, equivalently, mass of the doubly charged boson can be set 
infinitely large.) In Fig. 5 $M_{\Delta^{++}}=600$ GeV and one can see that
 these  curves look similar to those of Fig. 4.
Threshold behaviour  at $M_W \sim M_N$ can be explained by the change of sign
of propagator terms (diagrams 2 and 4):
$$
\Pi (p_1-p_4-p_5,M_N,\Gamma_N)= 
\frac{1}{M_W^2-M_N^2-2p_1 \cdot (p_4+p_5)+iM_N \Gamma_N}.
$$

In order to study the influence of $M_{\Delta^{++}}$ on the cross section
Fig. 6 shows the discovery limits of the reaction
$e^- \, \gamma \rightarrow  e^+ \,W_R^- \, W_R^-$
in the $M_N - M_{\Delta^{++}}$ plane for the
$\sqrt{s_{ee}}=$ 5 TeV and $M_{W_R}$= 1 TeV.
Final state $W_R$ are assumed to decay into
light quark jets (third generation excluded),
and the $4$-quark  efficiency is taken to be
85 \% and purity 80 \%, with the efficiency for
final $e^+$ 90 \%
\cite{Hemingway}.
The contour levels in the Figure are drawn
at 63 \%, 95 \% and  \mbox{99 \%}  probability of the reaction dicsovery
(1, 3 and 4.6 events per  year with the  anticipated luminosity).
The excluded region lies below the curves because of
lower limit on the Majorona mass of the neutrino.
One can see the threshold
 at $M_{\Delta^{++}}= 2 M_{W_R}$, it happens due to pole
in the $\Delta^{++}  $ propagator which gets involved
in the integration over final states.
Above this threshold, the reaction $e^- \, \gamma \rightarrow  e^+ \,W_R^- \, W_R^-$
is observable even for the very low value of Majorana neutrino mass,
however even below this threshold the reaction remains observable
for neutrino masses above 500 GeV. This means that effects of doubly charged
Higgs mass are not important for the reaction studied.
In further calculations we set $M_{\Delta^{++}}=600$ GeV
which keeps the $\Delta^{++}  $ propagator off-shell for realistic values of doubly
charged boson masses (and hence the reaction is studied in more conservative
regime, without possible $\Delta^{++}  $ pole enhancement ).

 It is also important to state here that  the process under study
takes place only due to the Majorana mass of the neutrinos,
in other words
all amplitudes presented in the Appendix are proportional to:
\begin{equation}
A^n \sim   M_{N} \cdot a^n(M_{N})
\end{equation}
and vanish for the case of massless neutrinos.
They can be generalized for the general case of 
right-handed neutrino mixings:
\begin{equation}
A^n \sim  \sum_i U_{ei}^2 \cdot M_{Ni} \cdot a^n(M_{Ni}).
\end{equation}

 From Figs. 3, 5
 and 6 it is possible
to conclude that for neutrino masses $M_N< M_{W_R}$
the cross section of the process  essentially decreases
and at some point $M_N \sim 500$ GeV the process becomes
invisible for the case of $M_{\Delta^{++}}< 2 M_{W_R}$.
If the latter condition does not hold the cross section 
still turns off
but at the lower values of $M_N$. 
 Hence, if the right-handed mass spectrum is such that
only one neutrino mass state is heavy enough to be 
`` visible'' (let us denote it as $m_{heavy}$),
the results presented may be generalized
for the process
$e^- \, \gamma \rightarrow  l^+ \,W_R \, W_R$
with muon or $\tau$ -lepton in the final state.
Defining in the standard way the effective neutrino masses:
$$
<m_{ik}>= \sum_j U_{ij} \cdot U_{kj} m_j
$$
one can treat all the  figures
as referring to arbitrary final lepton with the following changes:
\begin{equation}
\sigma(e^- \, \gamma \rightarrow  e^+ \,W_R^- \, W_R^- , M_N)
\rightarrow \frac{m^2_{heavy}}{<m_{el}>^2}
\sigma (e^- \, \gamma \rightarrow  l^+ \,W_R^- \, W_R^-, m_{heavy}) 
\end{equation}
$$
M_N \rightarrow m_{heavy}.
$$

 Finally, in Fig. 7 
are depicted the discovery limits for the studied process
in the $M_N - M_{W_R}$ plane for CLIC options 
 $\sqrt{s_{ee}}=$ 3 TeV {\bf (a)},  $\sqrt{s_{ee}}=$ 5 TeV {\bf (b)}
and $\sqrt{s_{ee}}=$ 8 TeV {\bf (c)}.
As indicated $M_{\Delta^{++}}$ is set at $600$ GeV.
The treatment of the final $W_R$ decays is the same as for Fig. 6.
The contour levels in the Figure are drawn
at 63 \%, 95 \% and \mbox{ 99 \%}  probability of the reaction discovery
(1, 3 and 4.6 events per  year with the anticipated luminosity).
The excluded region is above the curves.
It is possible to state, that the reaction
$e^- \, \gamma \rightarrow  e^+ \,W_R^- \, W_R^-$
will be observed for heavy Majorana neutrinos,
whose masses may essentially exceed straightforward discovery limit
($M_N > \sqrt{s_{ee}}  $) for reasonable values of right-handed charged bosons.
The corresponding lower limit on  $M_N$ increases with the increase
of charged boson mass, and the ``resonance-like'' behaviour of the 
contour-levels occurs due to interplay of $M_N$ and $M_{W_R}$
in the propagators of diagrams 2 and 4 (see comments to Figs. 4 and 5).

We do not present here the angular distributions of the final lepton since
--in complete accordance with \cite{Peressutti}-- the shape of these
distributions is non-universal and is governed by the interplay
between $\sqrt{s}$ and $M_N$. 

\section{Summary}

The LR model with heavy Majorana neutrinos is one of the most natural
and popular extensions of the SM. Observation of
right-handed gauge bosons, Majorana neutrinos and triplet Higgs bosons
 are necessary steps for the confirmation of this theory. 
 The reaction $e^- \, \gamma \rightarrow  e^+ \,W_R^- \, W_R^-$
can be discovered at CLIC
for a reasonable range of values of LR model parameters,
providing a good test of lepton number violation
in LR model.
 The reaction would be a
serious manifestation of a heavy Majorana neutrino.
  It will be   observable for realistic mass values of
right-handed bosons
in the neutrino mass range $M_N \sim \sqrt{s}$
 and even  well
above  limits for direct $N$ production ($M_N > \sqrt{s}$).
Discovery limits on $M_N$ depend on the corresponding value of
$M_{W_R}$ are very weakly dependent on
the doubly charged boson mass. However, extra heavy  doubly
charged Higgs ($M_{\Delta^{++}} > 2M_{W_R}$) can 
increase the probability of the reaction. If the right-handed
Majorana neutrino spectrum has only one  heavy eigenvalue,
all the results can be applied to the process 
$e^- \, \gamma \rightarrow  l^+ \,W_R \, W_R$ with the arbitrary
lepton in the final state. 

\acknowledgments

This research was supported in part by the Natural Sciences and Engineering 
Research Council of Canada and partially supported by
RFFI Grant 01-02-17152  (Russian Fund of Fundamental Investigations)
and by INTAS grant 2000-587 .
I would like to thank Prof. Pat Kalyniak for careful reading
of the manuscript and Profs. Steve Godfrey and
Richard Hemingway for useful discussions.

\section{Appendix}

We present here helicity amplitudes for Feynman diagrams 1-9
of Fig. 1. I use the following notations for particle momenta:
$p_1,\, p_2$ are the incident electron and photon moments correspondingly.
$p_3$ stands for outgoing positron (positively charged lepton),
$p_{45}, \, p_{67}$ are correspondingly, momenta of outgoing
right-handed bosons, where
$p_{ij}=p_i+p_j$ and $p_i, \, p_j$ are 4-momenta of massless fermions
 (see the CALCUL technique of massive gauge bosons representation),
$k$ is the photon CALCUL representation vector \cite{CALCUL}.
The gauge couplings are: $e$-electrical charge,
$g_R$ right-handed SU(2) gauge coupling. $M_N,\, \Gamma_N, \, M_W, \,\Gamma_W, \,
M_{\Delta}, \, \Gamma_{\Delta} $ denote the masses and widths of
right-handed Majorana neutrino, charged gauge boson and doubly charged
Higgs correspondingly. The propagator function is defined as follows:
$$\Pi (k,M,\Gamma)= \frac{1}{k^2-M^2+iM \Gamma}.$$
The upper index of the amplitudes denotes the number of 
the corresponding Feynman diagram, the lower index (L, R)
denotes the chirality of the incoming photon.
\begin{equation}
A^1_{\lambda}= \frac{i eg^2_R}{2}\cdot \frac{1}{\sqrt{4kp_2}}\cdot
\Pi (p_1+p_2,0,0) \cdot \Pi(p_1+p_2-p_{45}, M_N, \Gamma_N)
 \cdot \hat{ A^1_{\lambda}} 
+(45 \leftrightarrow 67), 
\end{equation}
$$
 \hat{ A^1_{L} }= 8M_N \cdot s(p_1,k)t(p_2,p_1)s(p_1,p_4)
                      t(p_5,p_7)s(p_6,p_3), 
$$
$$
\hat{ A^1_{R} }=   8M_N \cdot s(p_1,p_2)\left[t(k,p_1)s(p_1,p_4)
                     +t(k,p_2)s(p_2,p_4)\right]
                   t(p_5,p_7)s(p_6,p_3).  
 $$
\begin{equation}
A^2_{\lambda}= \frac{i eg^2_R}{2}\cdot \frac{1}{\sqrt{4kp_2}}\cdot
\Pi (p_3-p_2,0,0) \cdot \Pi(p_1-p_{45}, M_N, \Gamma_N)
 \cdot \hat{ A^2_{\lambda}} 
+(45 \leftrightarrow 67), 
\end{equation}
$$  
\hat{ A^2_{L}}=-8M_N s(p_1,p_4)t(p_5,p_7)
                       s(p_6,p_3)t(p_3,p_2)s(k,p_3),
$$ 
$$
 \hat{ A^2_{R}}=-8M_N s(p_1,p_4)t(p_5,p_7)
     \left[s(p_6,p_3)t(p_3,k)s(p_2,p_3)-s(p_6,p_2)t(p_2,k)s(p_2,p_3)
     \right].
$$
\begin{equation}
 A^3_{\lambda}= \frac{i eg^2_R}{2}\cdot \frac{1}{\sqrt{4kp_2}}\cdot
 \Pi(p_1+p_2-p_{45}, M_N, \Gamma_N) \cdot \Pi(p_{45}-p_2, M_W,0)
 \cdot \hat{ A^3_{\lambda}} 
+(45 \leftrightarrow 67), 
\end{equation}
$$
\hat{ A^3_{L}}=-8M_N s(p_6,p_3)
\left\{
s(p_1,p_2)t(p_2,p_7)s(p_4,k)t(p_2,p_5)- 
\right. 
$$ 
$$
\left.
s(p_1,p_4)t(p_5,p_7) \left[t(p_2,p_4)s(p_4,k)+t(p_2,p_5)s(p_5,k)\right]
-s(p_4,p_2)t(p_2,p_5)s(p_1,k)t(p_2,p_7)
\right\},
$$ 
$$
\hat{ A^3_{R}}=-8M_N s(p_6,p_3)
\left\{
s(p_1,p_2)t(p_2,p_7)s(p_4,p_2)t(k,p_5)-
\right. 
$$
$$ 
\left.
s(p_1,p_4)t(p_5,p_7) \left[s(p_2,p_4)t(p_4,k)+s(p_2,p_5)t(p_5,k)\right]
-s(p_4,p_2)t(p_2,p_5)s(p_1,p_2)t(k,p_7)
\right\}.
$$
\begin{equation}
 A^4_{\lambda}= \frac{i eg^2_R}{2}\cdot \frac{1}{\sqrt{4kp_2}}\cdot
 \Pi(p_1-p_{45}, M_N, \Gamma_N) \cdot \Pi(p_{67}-p_2, M_W,0)
 \cdot \hat{ A^4_{\lambda}} 
+(45 \leftrightarrow 67), 
\end{equation}
$$
\hat{ A^4_{L}}=-8M_N s(p_1,p_4)
\left\{-t(p_5,p_7)s(p_6,p_3)\left[t(p_2,p_6)s(p_6,k)+
                            t(p_2,p_7)s(p_7,k) \right]
\right. 
$$
$$
\left.
-s(p_6,p_2)t(p_2,p_7)t(p_5,p_2)s(k,p_3)+
t(p_5,p_2)s(p_2,p_3)s(p_6,k)t(p2,p7)
\right\}, 
$$
$$
\hat{ A^4_{R}}=-8M_N s(p_1,p_4)
\left\{-t(p_5,p_7)s(p_6,p_3)\left[s(p_2,p_6)t(p_6,k)+
                            s(p_2,p_7)t(p_7,k) \right]
\right. 
$$
$$
\left.
-s(p_6,p_2)t(p_2,p_7)t(p_5,k)s(p_2,p_3)+
t(p_5,p_2)s(p_2,p_3)s(p_6,p_2)t(k,p7)
\right\}.
$$ 
\begin{equation}
A^5_{\lambda}= \frac{i eg^2_R}{2}\cdot \frac{1}{\sqrt{4kp_2}}\cdot
 \Pi(p_1+p_2, 0, 0) \cdot \Pi(p_{45}+p_{67}, M_{\Delta},\Gamma_{\Delta})
 \cdot \hat{ A^5_{\lambda}},
\end{equation}
$$
\hat{ A^5_{L}}=8M_N s(p_1,k)t(p_2,p_1)s(p_1,p_3)s(p_4,p_6)t(p_7,p_5),
$$
$$ 
\hat{ A^5_{R}}=8M_N s(p_1,p_2)
\left[t(k,p_1)s(p_1,p_3)+t(k,p_2)s(p_2,p_3)
\right]
s(p_4,p_6)t(p_7,p_5).
$$
\begin{equation}
 A^6_{\lambda}= \frac{i eg^2_R}{2}\cdot \frac{1}{\sqrt{4kp_2}}\cdot
 \Pi(p_3-p_2, 0, 0) \cdot \Pi(p_{45}+p_{67}, M_{\Delta},\Gamma_{\Delta})
 \cdot \hat{ A^6_{\lambda}},
\end{equation}
$$
\hat{ A^6_{L}}=-8M_N s(p_4,p_6)t(p_7,p_5)s(p_1,p_3)t(p_3,p_2)s(k,p_3),
$$
$$
\hat{ A^6_{R}}=-8M_N s(p_4,p_6)t(p_7,p_5)
\left[s(p_1,p_3)t(p_3,k)-s(p_1,p_2)t(p_2,k) \right]s(p_2,p_3).
$$
\begin{equation}
 A^7_{\lambda}= \frac{i eg^2_R}{2}\cdot \frac{1}{\sqrt{4kp_2}}\cdot
 \Pi(p_{1}-p_{3}, M_{\Delta},\Gamma_{\Delta}) \cdot
\Pi(p_{45}-p_2, M_W,0)
 \cdot \hat{ A^7_{\lambda}}, 
\end{equation}
$$
\hat{ A^7_{L}}=-8M_N s(p_1,p_3) 
\left\{
s(p_4,k)t(p_2,p_5)s(p_6,p_2)t(p_2,p_7)-s(p_6,k)t(p_2,p_7)s(p_4,p_2)t(p_2,p_5)
\right. 
$$
$$
\left.
-\left[t(p_2,p_4)s(p_4,k)+t(p_2,p_5)s(p_5,k) \right] s(p_6,p_4)t(p_5,p_7)
\right\}, 
$$
$$
\hat{ A^7_{R}}=-8M_N s(p_1,p_3) 
\left\{
s(p_4,p_2)t(k,p_5)s(p_6,p_2)t(p_2,p_7)-s(p_6,p_2)t(k,p_7)s(p_4,p_2)t(p_2,p_5)
\right. 
$$
$$
\left.
-\left[s(p_2,p_4)t(p_4,k)+s(p_2,p_5)t(p_5,k) \right] s(p_6,p_4)t(p_5,p_7)
\right\}. 
$$
\begin{equation}
A^8_{\lambda}=A^7_{\lambda}(p_4 \leftrightarrow p_6, p_5 \leftrightarrow p_7).
\end{equation}
\begin{equation}
 A^9_{\lambda}= \frac{i eg^2_R}{2}\cdot \frac{1}{\sqrt{4kp_2}}\cdot
 \Pi(p_1-p_3, M_{\Delta}, \Gamma_{\Delta}) \cdot 
\Pi(p_{45}+p_{67}, M_{\Delta},\Gamma_{\Delta})
 \cdot \hat{ A^9_{\lambda}}, 
\end{equation}
$$
\hat{A^9_{L}}=16 M_N
\left[t(p_2,p_1)s(p_1,k)-t(p_2,p_3)s(p_3,k) \right]
s(p_1,p_3)s(p_4,p_6)t(p_7,p_5),
$$
$$
\hat{A^9_{R}}=16 M_N
\left[s(p_2,p_1)t(p_1,k)-s(p_2,p_3)t(p_3,k) \right]
s(p_1,p_3)s(p_4,p_6)t(p_7,p_5).
$$

\newpage
\begin{figure}
\hspace{-2cm}
           {\epsfig{file=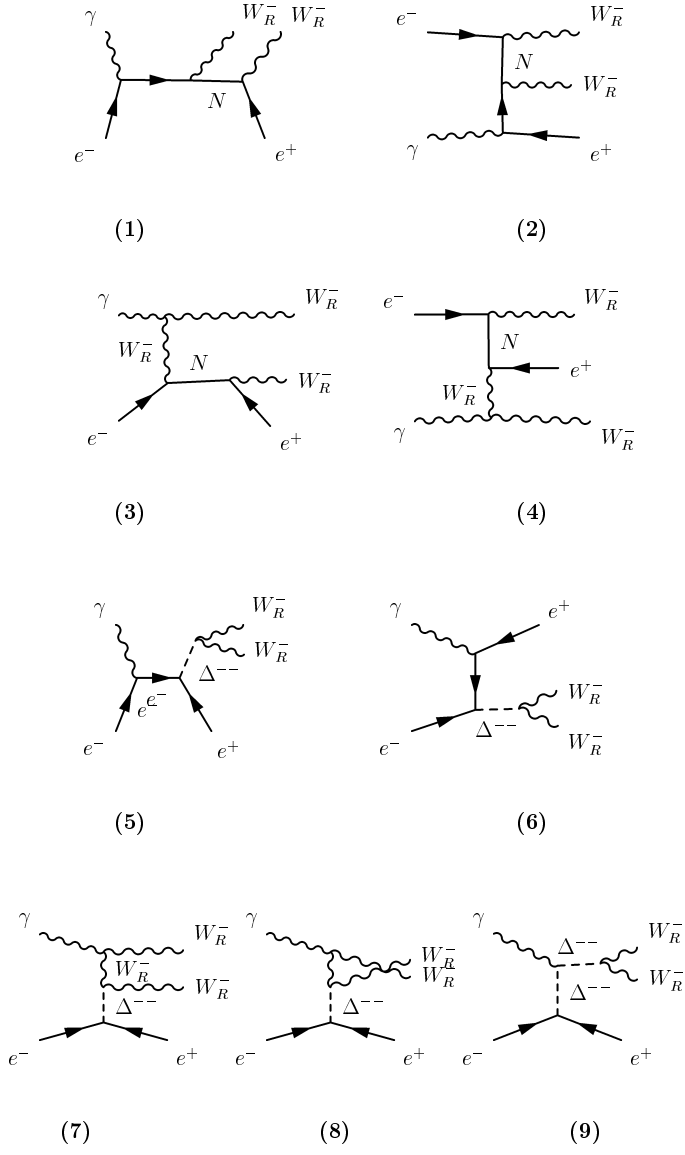,width=6.0in}
           }
\vspace{20pt}
\caption{The Feynman diagrams for the $e^- \, \gamma \rightarrow  e^+ \,W_R \, W_R$
process.}
\label{Fig1}
\end{figure}

\newpage
\begin{figure}
\centerline{\epsfig{file=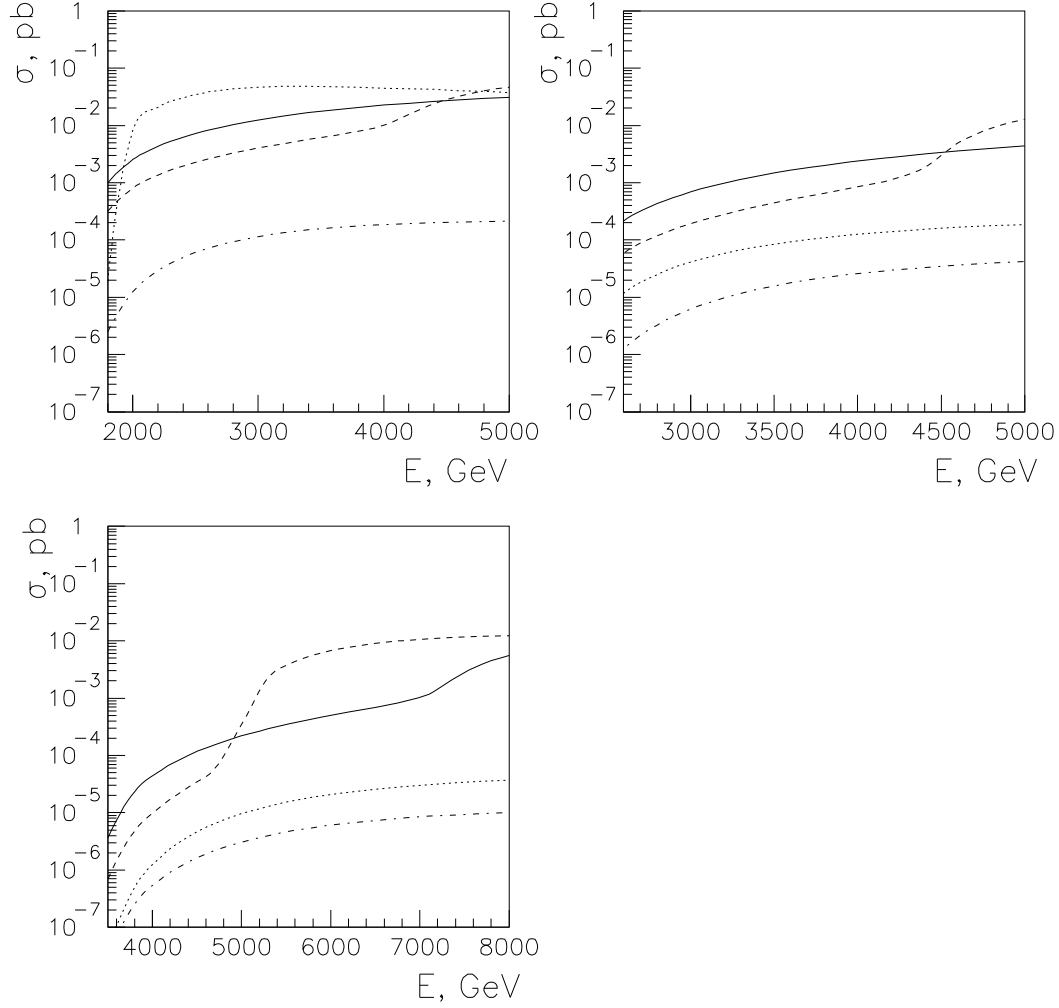,width=6.0in}
           }
\vspace{20pt}
\caption{ 
The cross section $\sigma (e^- \, \gamma  \rightarrow e^+ \,
 W_R^- \, W_R^-)$  as a function of $\sqrt{s_{ee}}$.
The mass of the doubly 
charged Higgs is $M_{\Delta_{++}}=600$ GeV. 
$M_{W_R}=$ 700, 1000, 1500 GeV for
upper left, upper right and lower paints correspondingly. 
In all three cases solid line is for neutrino mass
$M_N=5$ TeV, long- dashed line  for $M_N=3$ TeV, short-dashed
for $M_N=1$ TeV, dashed-dotted line for $M_N=500$ GeV.
 }
\label{Fig2}
\end{figure}

\newpage
\begin{figure}
\centerline{ \epsfig{file=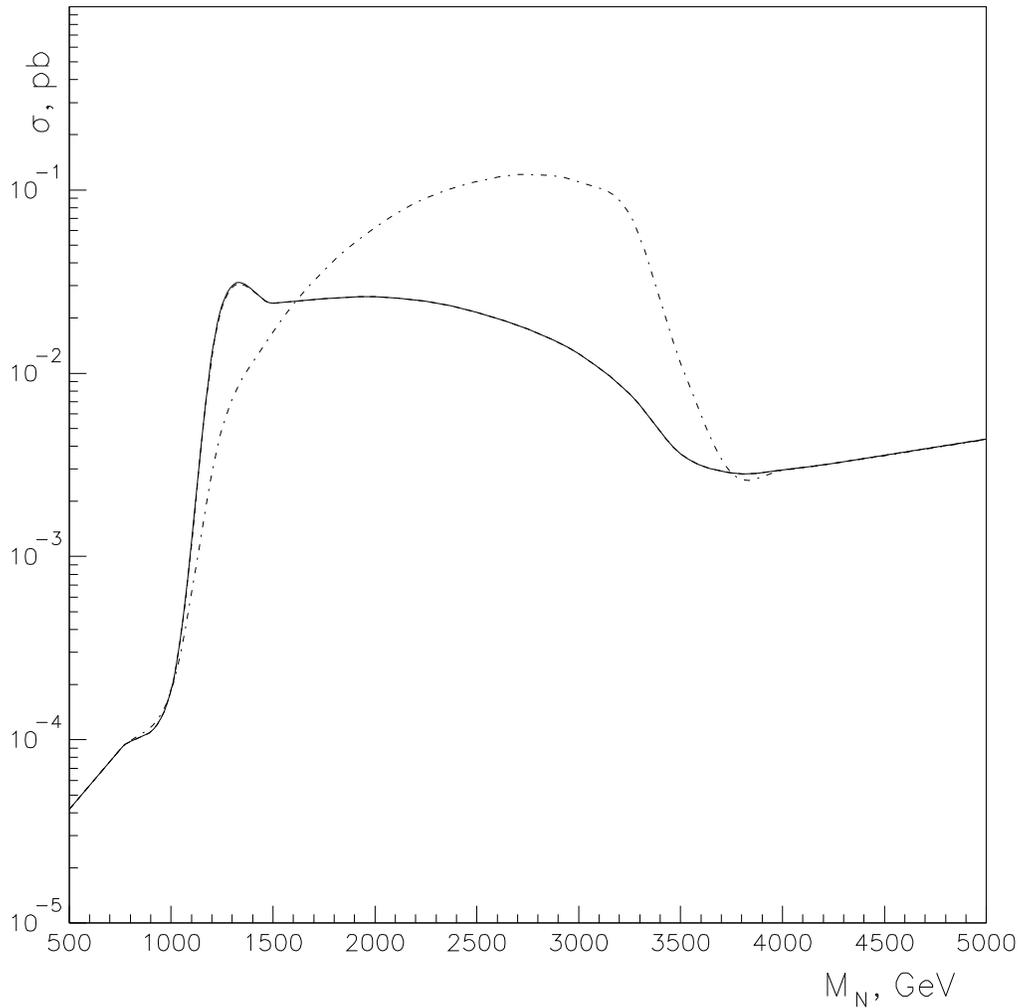,width=6.0in}
           }	   
\vspace{20pt}
\caption{
The cross section $\sigma (e^- \, \gamma  \rightarrow e^+ \,
 W_R^- \, W_R^-)$  as a function of right-handed Majorana neutrino mass
for $M_{W_R}=1$ TeV, $\sqrt{s_{ee}}=$ 5 TeV. The dash-dotted line is
for constant neutrino and doubly charged higgs widths
$\Gamma_N=\Gamma_{\Delta^{++}}=10$ GeV,
 solid line is for realistic, mass dependent widths. Dashed line
( completely coinciding with solid one)
is for $ \Gamma_{\Delta^{++}}=10$ GeV   and realistic $\Gamma_N$.    
}
\label{Fig3}
\end{figure}

\newpage
\begin{figure}
\centerline{\epsfig{file=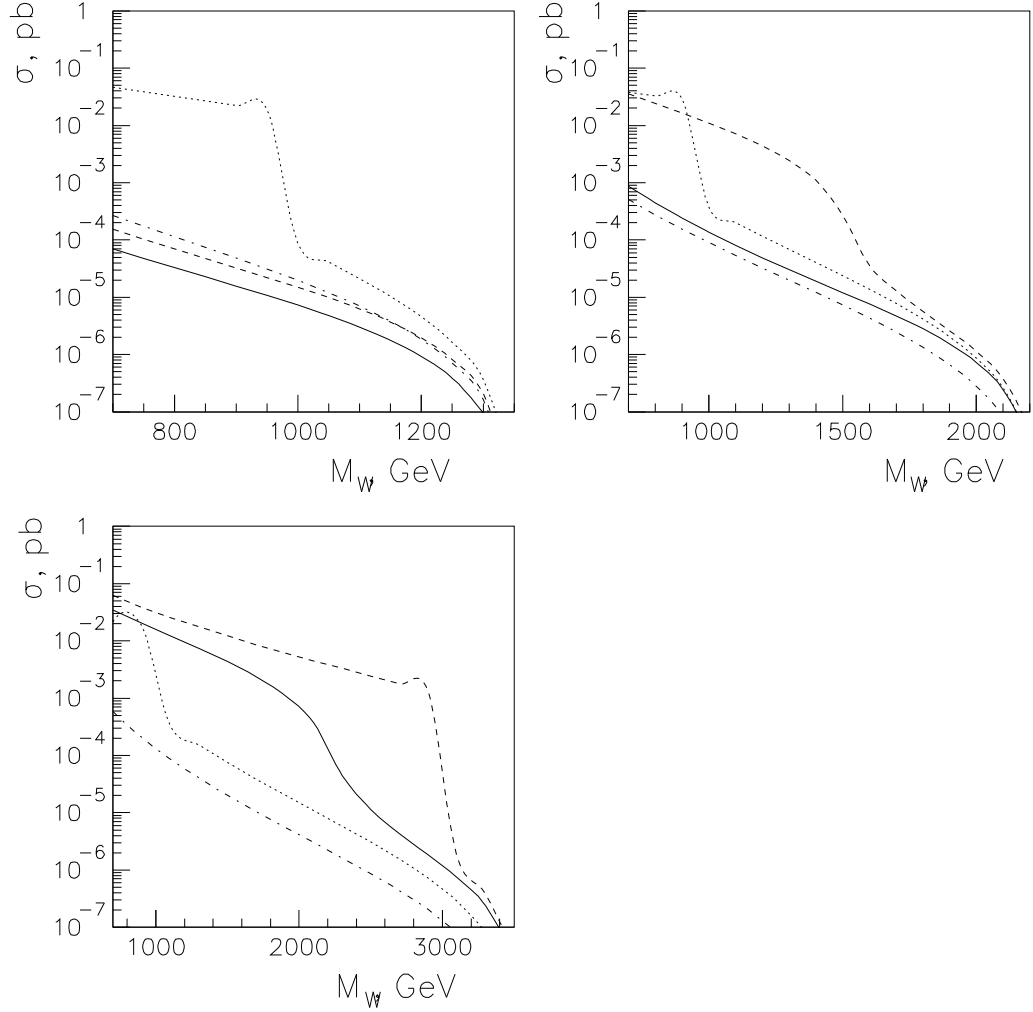,width=6.0in}
           }
\vspace{20pt}
\caption{
The cross section $\sigma (e^- \, \gamma  \rightarrow e^+ \,
 W_R^- \, W_R^-)$  as a function of $M_{W_R}$. Diagrams with doubly 
charged Higgs boson are neglected. $\sqrt{s_{ee}}=$ 3,5,8 TeV for
upper left, upper right and lower paints correspondingly. 
In all three cases the solid line is for neutrino mass
$M_N=5$ TeV, the long - dashed line  for $M_N=3$ TeV, the short-dashed
for $M_N=1$ TeV, the dashed-dotted line for $M_N=500$ GeV.
   }
\label{Fig4}
\end{figure}

\newpage
\begin{figure}
\centerline{\epsfig{file=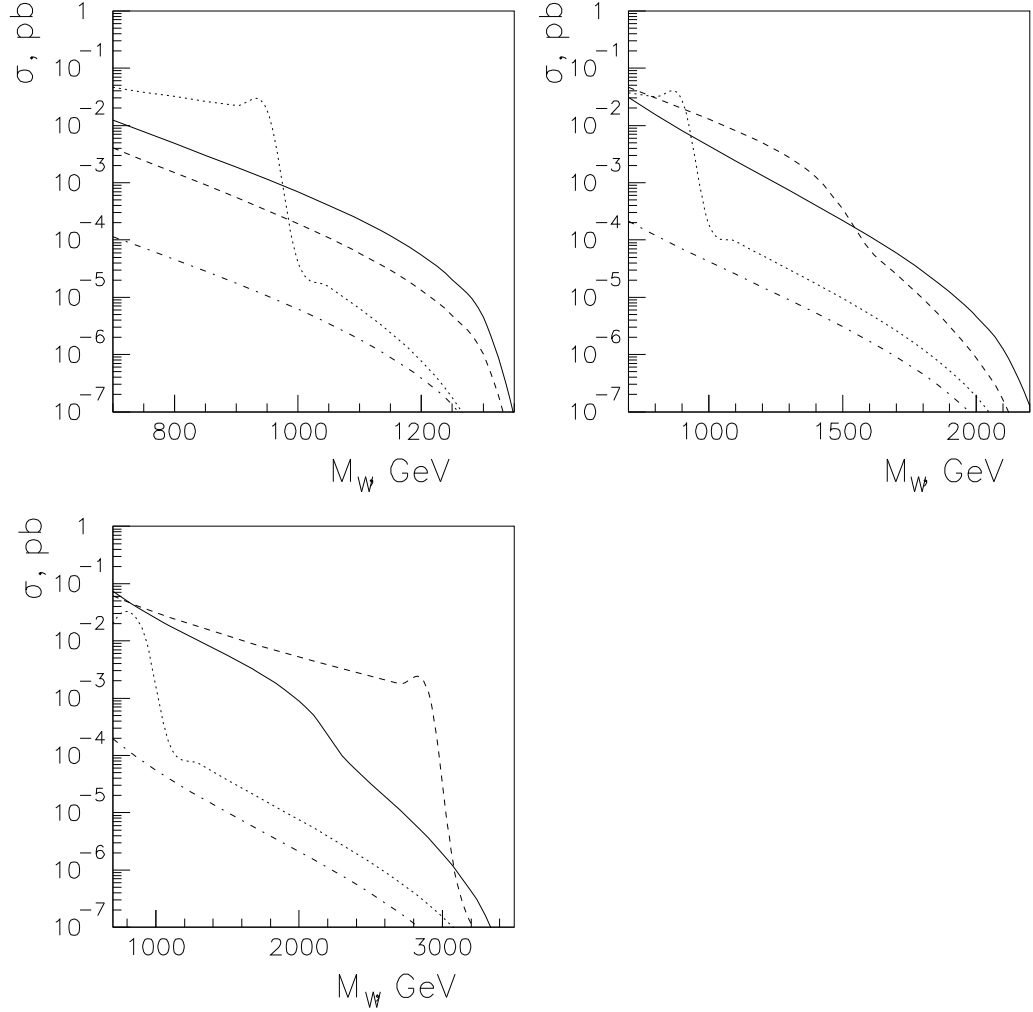,width=6.0in}
           }
\vspace{20pt}
\caption{ 
The cross section $\sigma (e^- \, \gamma  \rightarrow e^+ \,
 W_R^- \, W_R^-)$  as a function of $M_{W_R}$. The mass of the doubly 
charged Higgs is $M_{\Delta_{++}}=600$ GeV. 
$\sqrt{s_{ee}}=$ 3,5,8 TeV for
upper left, upper right and lower paints correspondingly. 
In all three cases the solid line is for neutrino mass
$M_N=5$ TeV, the long-dashed line  for $M_N=3$ TeV, the short-dashed
for $M_N=1$ TeV, the dashed-dotted line for $M_N=500$ GeV.
 }
\label{Fig5}
\end{figure}

\newpage
\begin{figure}
\centerline{\epsfig{file=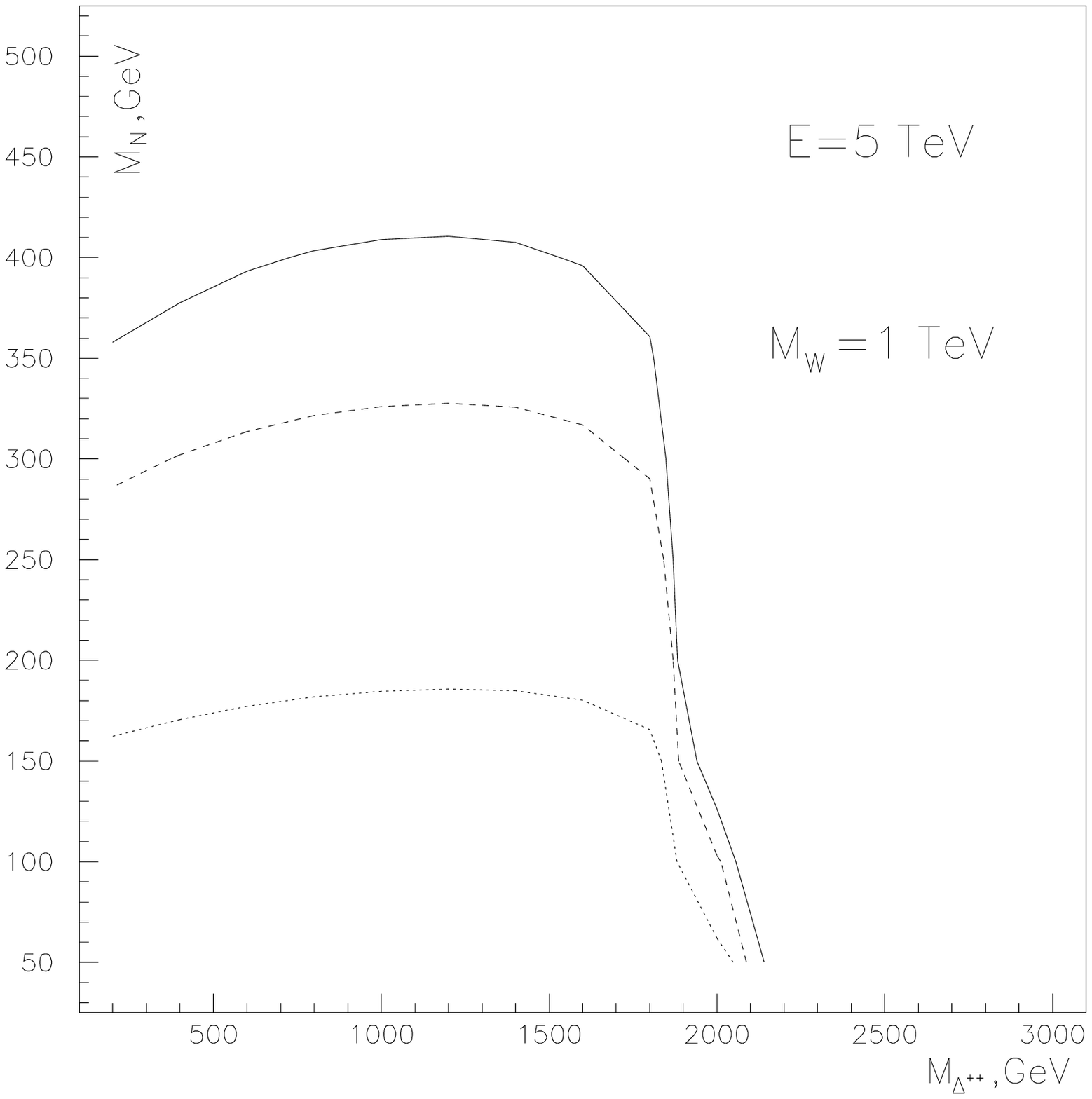,width=6.0in}
           }
\vspace{20pt}
\caption{The contour levels in the $M_N$--$M_{\Delta^{++}}$
 plane that correspond to 99 \% (solid line),
  95 \% (long-dashed line) and 63 \% (short-dashed line) probability
of the discovery level (4.6, 3 and 1 event) for
$\sqrt{s}=5$ TeV, $M_{W_R}=1$ TeV.}
\label{Fig6}
\end{figure}

\newpage
\begin{figure}
\centerline{\epsfig{file=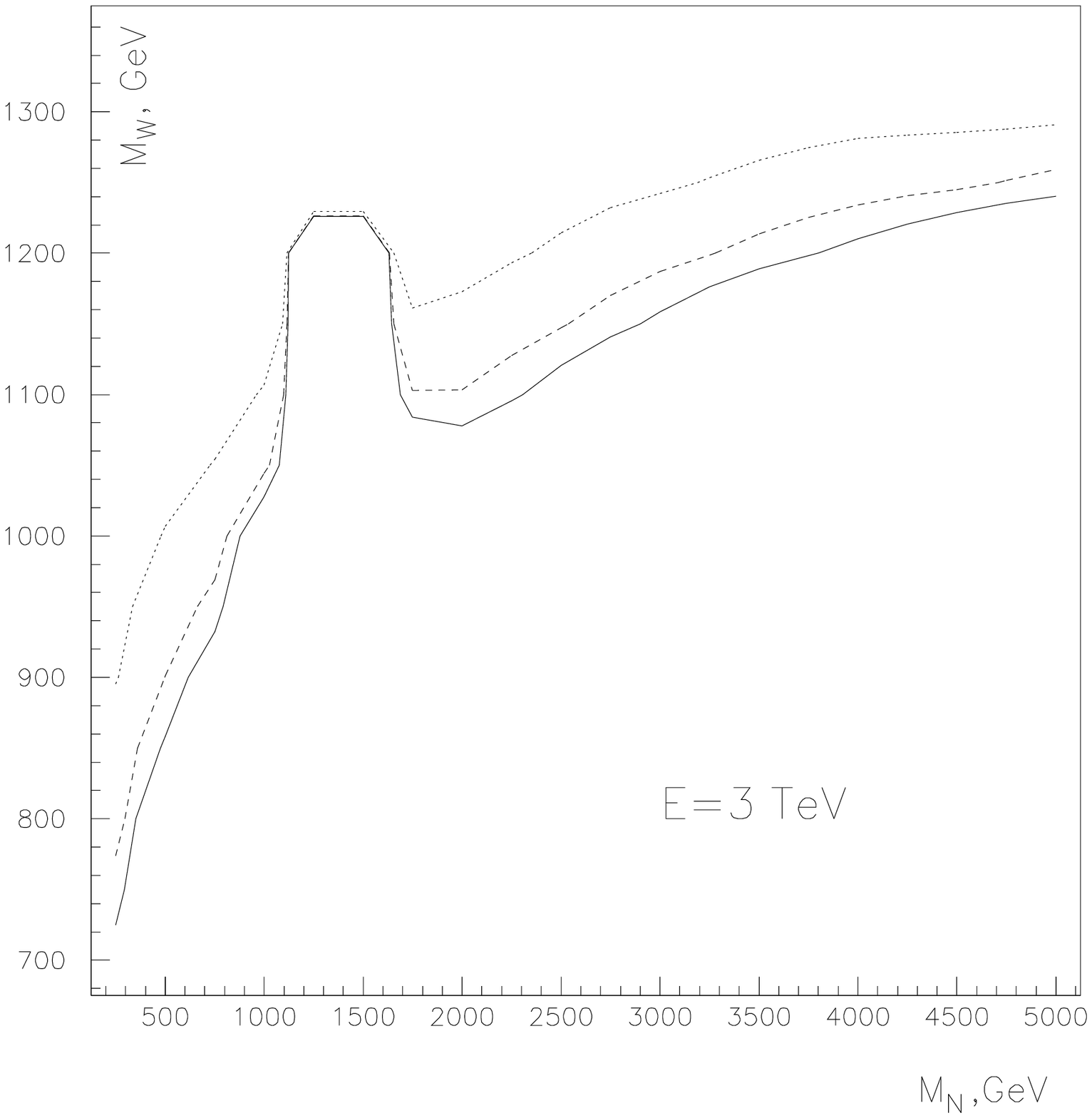,width=3.0in}
            \epsfig{file=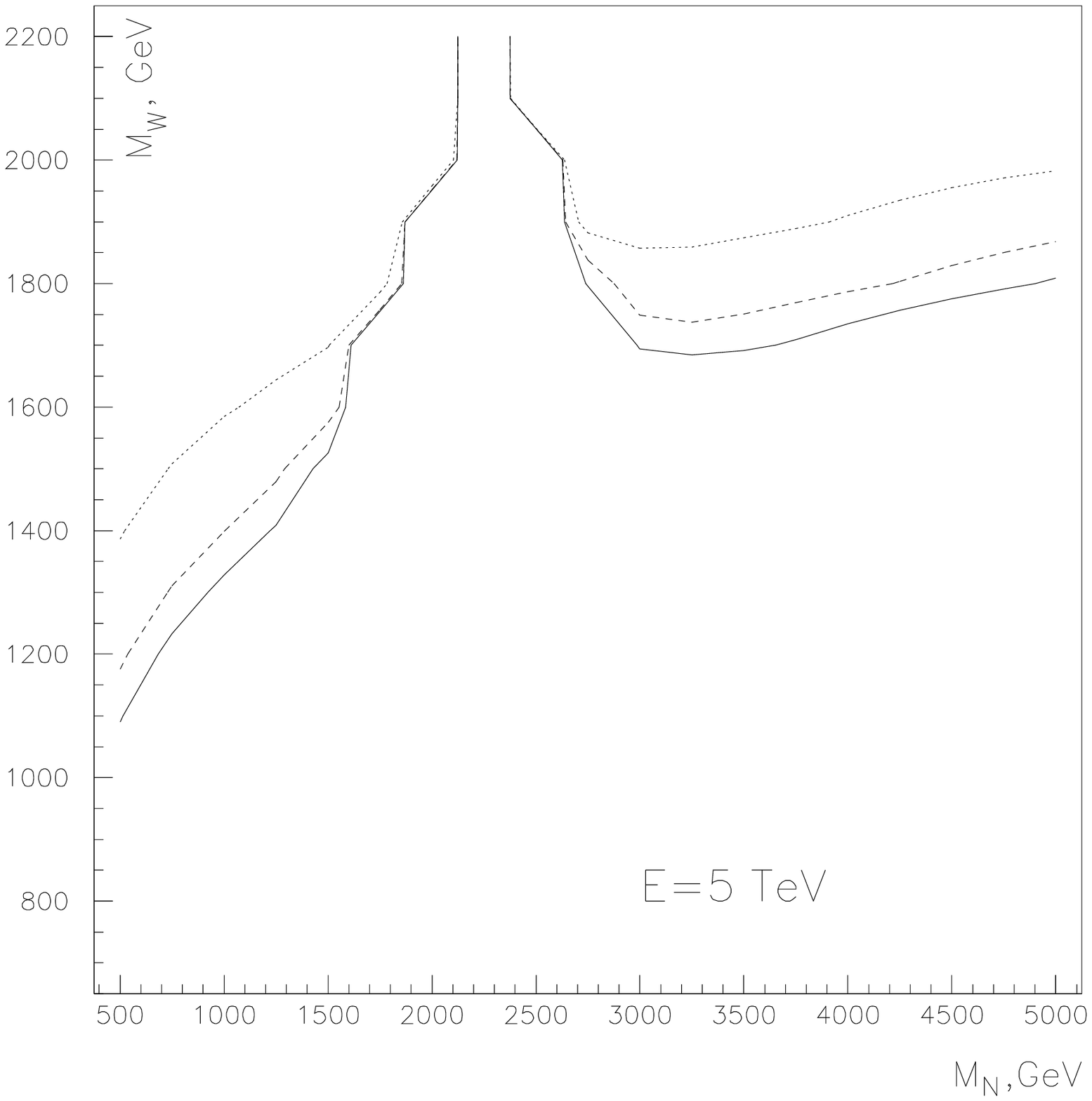,width=3.0in}
           }
\centerline{ \bf (a) \ \ \ \ \hspace{5cm} \ \ \ \ \ \ \ (b)}
\centerline{\epsfig{file=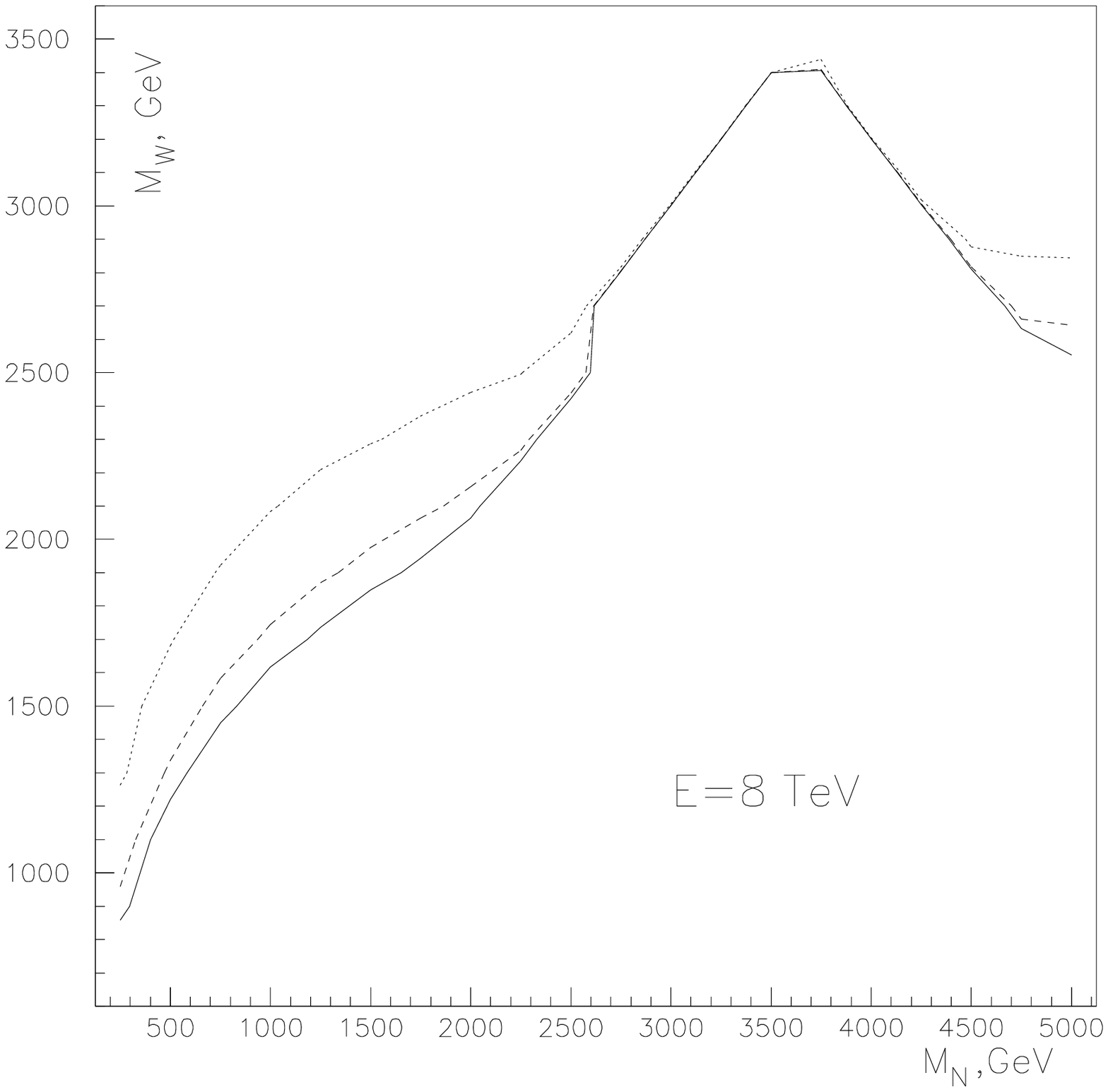,width=3.0in}
           }
\centerline{ \bf (c)}
\vspace{20pt}
\caption{The contour levels in the $M_N$--$M_{W_R}$ plane that 
correspond to 99 \% (solid line), 95 \%
(long-dashed line) and 63 \% (short-dashed line) probability
of the discovery level (4.6, 3 and 1 event) for 
$\sqrt{s_{ee}}=$ 3 {\bf (a)}, 5 {\bf (b)}, 8 {\bf (c)} TeV.
$M_{\Delta^{}++}=$ 600 GeV. 
}
\label{Fig8}
\end{figure}

\end{document}